%% file: main.tex
\documentclass[11pt,letterpaper]{article}
\ifdefined\pdfminorversion\pdfminorversion=7\fi

\usepackage[margin=1in]{geometry}
\usepackage[T1]{fontenc}
\usepackage{lmodern}
\usepackage{etoolbox}
\usepackage{amsmath,amssymb}
\usepackage{graphicx}
\usepackage{array,booktabs,tabularx}
\usepackage{xcolor}
\usepackage{xspace}
\usepackage[labelfont=bf]{caption}
\usepackage[authoryear,round]{natbib}
\usepackage{microtype}
\usepackage{placeins}
\usepackage{hyperref}
\AtEndPreamble{\hypersetup{
  hidelinks,
  pdftitle={ModelLakeFishing: Efficient Retrieval over Million-Scale Model Lakes},
  pdfauthor={Xiaoyang Liu, Zhengyuan Dong, Renee J. Miller}
}}


\newcommand{\ours}{\textsc{ModelLakeFishing}\xspace}
\newcommand{\exact}{\textsc{Exact}\xspace}
\newcommand{\exhaustive}{\textsc{Exhaustive}\xspace}

\newcommand{\repositoryURL}{https://github.com/ERHUTUZI123/ModelLakeFishing}
\newcommand{\codeavailability}{%
  \ifdefempty{\repositoryURL}{}{%
    \footnote{The code is available at
      \expandafter\url\expandafter{\repositoryURL}.}}}
\AtBeginDocument{\ifdefempty{\repositoryURL}{%
  \PackageWarningNoLine{arxiv-preprint}{Public GitHub URL is not set; fill repositoryURL before submission}}{}}

\title{\ours: Efficient Retrieval over Million-Scale Model Lakes}
\author{%
  Xiaoyang Liu\quad Zhengyuan Dong\quad Ren\'ee J. Miller\\[0.5em]
  University of Waterloo\\[0.3em]
  \small\texttt{\{x98liu,zhengyuan.dong,rjmiller\}@uwaterloo.ca}}
\date{}

\begin{document}
\maketitle

\input{src/abstract}
\input{src/introduction}
\input{src/related}

\input{src/methodology}
\input{src/experiment}
\input{src/conclusion}

\FloatBarrier
\bibliographystyle{plainnat}
\bibliography{bib/topic-datalake,bib/topic-modellake,bib/topic-others,bib/topic-ourgroup,bib/topic-llm,bib/topic-scitable}

\clearpage
\appendix
\input{src/appendix}

\end{document}

%% file: src/abstract.tex
\begin{abstract}
Open model lakes may contain millions of reusable models, making 
it costly to identify suitable models for a new dataset.
This setting requires learning indexable representations of models from heterogeneous evidence, %
including model and dataset metadata, lineage, and historical evaluations.
An important challenge is that this heterogeneous evidence is sparse and concentrated around popular models and benchmarks.
We present ModelLakeFishing, a model-retrieval framework for queries that specify a target dataset, a prediction task, and an evaluation metric.
It consolidates model lake metadata and historical evaluations into a model–dataset–task evidence graph, learns model and query embeddings with a relation-aware graph encoder, and indexes model embeddings with Hierarchical Navigable Small World (HNSW) search. At query time, HNSW retrieves 1k candidates without 
scoring every model in the lake.
A task-level prior computed only from training-side evidence then reranks the candidates for the requested metric and returns the top 10 models.
We evaluate the framework on a lake containing 3,016,439 models and 247,803 observed model–dataset performance pairs using three root-aware splits in which test performance edges are held out from representation learning and retrieval.
ModelLakeFishing achieves a mean eligible-query $\texttt{gold@10}$ of 0.2968, recovering the observed-best model among the ten returned models for 29.68\% of eligible queries.
This retains 93.47\% of the $\texttt{gold@10}$ achieved by an exhaustive baseline that applies the same scoring and reranking procedure to every model in the lake.
Given precomputed query embeddings, candidate retrieval and reranking take 0.747~ms at the median and 1.102~ms at the 95th percentile.
These results show that sparse relational evidence can be distilled into an indexable representation space for efficient retrieval over million-model lakes.\codeavailability
\end{abstract}

%% file: src/introduction.tex
\section{Introduction}~\label{sec:intro}

\noindent\textbf{Recommendation over a large-scale model lake needs candidate discovery.}
Public model hubs now resemble \emph{model lakes}: large collections of reusable models accompanied by model cards, configurations, lineage, and evaluation metadata~\citep{DBLP:conf/edbt/PalBM25, DBLP:journals/corr/abs-2512-16106}.
Hugging Face alone hosts more than three million candidate models across diverse tasks, architectures, and modalities.
For a target dataset, trying every model is impractical, while computing a separate query-specific relevance score for every model still requires traversing the full lake.
The immediate problem is therefore to find a small, useful shortlist before any candidate is downloaded or run.

\noindent\textbf{Prior methods use different selection signals.}
Existing model-selection methods reduce the cost of assessing candidates in different ways.
Feature-based transferability estimators such as LEEP~\citep{nguyen2020leep}, LogME~\citep{you2021logme}, NLEEP~\citep{li2021nleep}, and H-score~\citep{bao2019hscore} avoid full fine-tuning yet still execute each candidate on target-dataset samples.
Other approaches use repository information to guide selection: TransferGraph uses learned graph representations in a performance regressor~\citep{li2024model}; ModelLens combines metadata and evaluation records in a task- and metric-conditioned interaction scorer~\citep{cai2026modellens}; and HuggingR$^{4}$ retrieves candidates from model-card representations for natural-language requests before iterative refinement~\citep{ma2025huggingr}.
Together, these works demonstrate interaction-based ranking and metadata-based retrieval, leaving open how sparse model--dataset evidence can support indexed candidate generation for dataset--task queries over the full lake.

\begin{figure}[htbp]
    \centering
    \includegraphics[width=0.8\linewidth]{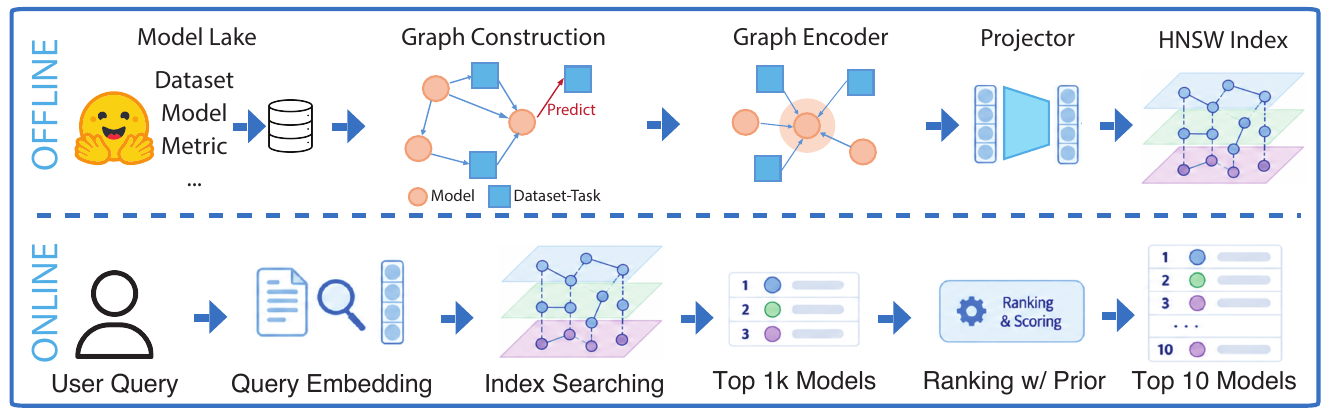}
    \caption{\textsc{ModelLakeFishing} overview. Everything above the dashed line is computed once, offline; everything below runs per query. The offline pipeline turns heterogeneous hub metadata and sparse performance evidence into a shared model--dataset--task embedding space and an HNSW index.}
    \label{fig:pipeline}
\end{figure}

\noindent\textbf{Relation-aware learning makes evidence searchable.}
Historical performance records cover only a small fraction of the model lake, so direct evaluations alone cannot support candidate discovery throughout the collection.
Model and dataset metadata provide descriptions that can be crawled and enriched~\citep{li2023macaroni}; observed performance, dataset similarity, and model lineage provide complementary relations with different meanings.
\ours\ brings these signals together in a heterogeneous model--dataset evidence graph and trains a relation-specific GraphSAGE encoder~\citep{hamilton2017graphsage} to place models and dataset--task queries in a shared representation space.
Unlike TransferGraph's downstream query--model performance regressor~\citep{li2024model}, our retrieval score is a direct inner product between independently computed representations.
This lets us index model representations offline with HNSW~\citep{DBLP:journals/pami/MalkovY20}, retrieve a bounded pool for a dataset--task query, and apply a historical same-task prior only when reranking that pool.
Figure~\ref{fig:pipeline} summarizes this offline and query-time division.

\noindent\textbf{A shortlist retains strong candidates.}
We assess candidate discovery by withholding performance records for entire dataset roots (groups of related configurations of the same dataset)%
 and asking whether the returned top ten include the best model \emph{observed in those held-out records}.
This observed-best recovery measure, \texttt{gold\@10}, is compared with a full-lake reference that uses the same learned representations and reranking score but considers every model.
Indexed retrieval retains most of the reference's recovery while limiting retrieval and reranking to a bounded candidate pool.
The lookup timing begins with a precomputed query representation, separating the cost of candidate discovery from query encoding.

\paragraph{Contributions.}
\noindent\textbf{(1) Candidate discovery through indexable representations.}
We learn model and dataset--task representations from heterogeneous model--dataset evidence so that models across the full lake can be indexed once and retrieved without candidate-wise execution or query-conditioned scoring over every model.
The retrieved candidate pool is then reranked using historical task evidence.

\noindent\textbf{(2) A full-lake evidence graph and leakage-safe evaluation protocol.}
We construct and audit a multi-million-model evidence graph and split performance records by dataset root %
into training, validation, and test sets.
Test performance records are used only for evaluation, not for representation learning or 
candidate discovery.

\noindent\textbf{(3) Quality and efficiency at three-million-model scale.}
Across three dataset-root partitions, the indexed retrieval (candidate discovery) and reranking system achieves a mean \texttt{gold\@10} of 0.2968 and retains 93.47\% of the \texttt{gold\@10} achieved by 
exhaustively scoring and reranking every model with the same representations and task prior. %
With query embeddings precomputed, candidate retrieval and reranking have a median latency of 0.747~ms and a 95th-percentile latency of 1.102~ms, with both values averaged across partitions. This compares with an average latency of 5.73~ms for the full-lake exhaustive scoring.

%% file: src/related.tex
\section{Related Work}~\label{sec:related}

\paragraph{Model lakes.}
Model management systems record experimental artifacts, versions, and intermediate results so that individual models can be reproduced, compared, or diagnosed~\citep{MISTIQUE2018,mei2022model}. Provenance research studies derivation and reuse across model versions~\citep{mu2024model}. A model lake extends this scope from a controlled collection to reusable models, datasets, metadata, lineage, and evaluation evidence at repository scale~\citep{DBLP:conf/edbt/PalBM25,garouani2024model, DBLP:journals/corr/abs-2512-16106}. For the massive model lakes we are considering, management and discovery are coupled since exhaustively considering every model for a tasks is infeasible. 
A useful model must be located before its artifacts can be inspected or evaluated, and evidence may be distributed across metadata, related repositories (for example code repositories), and historical evaluations. Public model lakes expose model cards and searchable repository metadata~\citep{modelcardpaper}; Macaroni crawls and enriches metadata from public model zoos~\citep{li2023macaroni}. Model-card documentation is nevertheless incomplete and uneven~\citep{liang2024whats}, limiting retrieval from lexical descriptions or normalized fields alone. 
An alternative to using documentation for search is to use behavior-based search (find models with a similar behavior)~\citep{lu2023content} or multimodal, meta-learning-based retrieval~\citep{li2023automrm}. 
These methods extend what a query can express, but the query itself still drives retrieval. 
Our setting is different: the evidence that makes a model retrievable is not only in the query but also in the lake's own relational structure — links among models, dataset–task pairs, observed performance, and model lineage, evidence that is sparse but structured. We ask how these relations can be distilled into an indexable search space over millions of candidates.

\paragraph{Model recommendation.}
Classical model selection ranks candidates through downstream validation or fine-tuning, but evaluating every model becomes prohibitive as the size of the model lake grows~\citep{ding2018modelselection}. Transferability estimators such as LEEP~\citep{nguyen2020leep}, LogME~\citep{you2021logme}, NLEEP~\citep{li2021nleep}, and H-score~\citep{bao2019hscore} avoid the expense of fully fine-tuning each candidate model on the target dataset, but still require running each candidate forward on target-dataset samples to extract its outputs or feature statistics.
They therefore choose within an available pool, whereas our setting requires a shortlist from the full lake before models are downloaded or run. HuggingR$^{4}$ frames repository-scale selection as iterative reasoning, retrieval, refinement, and reflection~\citep{ma2025huggingr}. 
ModelSearch uses structural similarity between model-card tables to recommend models that are both relevant to the user's request and diverse~\citep{dong2026diversed}.
Historical model--dataset interactions provide a complementary basis for recommendation without candidate execution: TransferGraph learns from a graph of models, datasets, similarities, and observed performance~\citep{li2024model}; ModelLens combines heterogeneous metadata with task- and metric-conditioned interaction evidence at larger scale~\citep{cai2026modellens}. Both demonstrate that relational history improves recommendation, but their reported final relevance functions score individual query--candidate pairs rather than expose a query-independent model index. We ask how relational supervision can instead produce an ANN-searchable space over millions of models.

\paragraph{Large-scale indexing.}
Learned retrieval makes search indexable by encoding queries and candidates independently, then comparing their vectors with a fixed similarity function. Sentence-BERT makes candidate representations reusable for similarity search~\citep{reimers2019sentence}; DPR encodes the corpus offline and uses approximate nearest-neighbor search to obtain a bounded candidate set for each online query~\citep{karpukhin2020dense}. This factorization is the interface between representation learning and indexing: unlike a joint query--candidate scorer, it lets candidate vectors be computed once rather than rescored across the full collection. Approximate nearest-neighbor systems reduce the cost of searching these spaces. HNSW provides high-recall search without an exhaustive scan in the typical query path~\citep{DBLP:journals/pami/MalkovY20}; Faiss provides complementary exact and approximate indexes, vector compression, and hardware-oriented implementations for large embedding collections~\citep{douze2024faiss}. These methods assume relevance has already been expressed in a searchable vector space. Our problem lies at that preceding boundary: sparse model--dataset observations and heterogeneous graph relations do not directly define indexable model vectors. \ours uses inductive graph representation learning~\citep{hamilton2017graphsage} to produce query-independent model embeddings and compatible dataset--task embeddings, indexes models once with HNSW, and reserves query-conditioned task evidence for reranking only the retrieved set.

%% file: src/methodology.tex
\section{Methodology}\label{sec:method}

\subsection{Problem Formulation and Overview}
\label{sec:method-overview}

\noindent\textbf{Candidate Discovery Task.}
Let $\mathcal D$ denote the dataset--task query nodes in the evidence graph.
Each query $d=(\operatorname{normalize}(\mathrm{dataset}),t)\in\mathcal D$ identifies a canonicalized dataset and a prediction task $t$
Different tasks on the same dataset can therefore induce different model rankings.
The objective is to return $k$ models from the candidate model lake $\mathcal M$ that recover strong candidates.
We judge strong candidates to be candidates that are ranked high in held-out historical evaluations.
Table~\ref{tab:method-notation} summarizes the notation used.

\begin{table*}[!htbp]
  \centering
  \small
  \caption{Notations.}
  \label{tab:method-notation}
  \renewcommand{\arraystretch}{1.12}
  \begin{tabularx}{\textwidth}{@{}lX@{\hspace{1.5em}}lX@{}}
    \toprule
    Notation & Description & Notation & Description \\
    \midrule
    $\mathcal M$ & Model lake.
      & $z_d,\;z_m$ & Query and model embeddings. \\
    $N=|\mathcal M|$ & Model lake size.
      & $s_\theta(d,m)$ & Retrieval score. \\
    $\mathcal D$ & Dataset--task nodes.
      & $\widehat{\mathcal P}_K(d)$ & Top-$K$ retrieved models. \\
    $d$ & Query node.
      & $p_t(m)$ & Same-task prior. \\
    $y_{dm}$ & Observed performance for $(d,m)$.
      & $r(d,m)$ & Final reranking score. \\
    \bottomrule
  \end{tabularx}
\end{table*}
\FloatBarrier

\noindent\textbf{Evidence Graph Construction.}
Historical evaluation records report how particular models performed on particular dataset--task pairs, but cover only a small part of the model lake.
We organize these records into a graph with model nodes and dataset--task nodes. Each recorded evaluation connects a model to the corresponding dataset--task node;
dataset simialrity and model lineage provide the other relations.
This graph supplies the context from which we learn model and query representations and is described in Section~\ref{sec:evidence-corpus}

\noindent\textbf{Learning and lookup are separate.}
\ours constructs an evidence graph and learns unit-normalized model and query representations offline.
At query time, it retrieves a bounded candidate pool from an index of model representations and reranks that pool using a same-task historical prior.
We collect historical model evaluation results from model cards and
existing evaluation collections. Each record identifies a model $m$,
a dataset--task pair $d$, and a measured performance score of $m$ on $d$.
After preprocessing, the score becomes $y_{dm}$, and the corresponding
pair defines a performance edge. We partition these records by dataset
into training, validation, and test sets. The nodes, metadata,
dataset-similarity edges, and model-lineage edges remain the same.
During training, 70\% of the training performance edges are included
in the message-passing graph. The remaining 30\% are held out of
the graph, and their scores provide supervision for the ranking loss.
Validation and test performance edges are excluded from the training
graph, and their scores are not used for training.
During testing, we measure recommendation quality on the test set.
Using the trained parameters, we compute model and query representations
with all training and validation performance edges as graph context,
then retrieve and rerank models for each test query.
Test performance edges are held out of this graph in both directions;
their recorded scores provide the ground truth for evaluating
the returned rankings.

\subsection{Evidence Graph Construction}
\label{sec:evidence-corpus}\label{sec:evidence-graph}

\noindent\textbf{Node:  Sparse records supply supervision.}
The evidence corpus combines Hugging Face crawls dated August 18, 2026 with five historical evaluation graphs.
We canonicalize model IDs and attach dataset cards only through exact or unambiguous repository matches; otherwise, the dataset name and task recorded in the evaluation record provide the description.  %
We create a node for every model in the crawl or referenced in the crawl. 
The referenced models include 12,680 absent from the crawl but unambiguously referenced crawled model or dataset cards. 
We assign each node a fixed integer ID and use it consistently for its
feature row, graph-edge endpoints, and learned embedding; model IDs also label entries in the search index.

We create a dataset--task node only when that pair appears in the evaluation records, whether or not a dataset card is available.
We group performance observations by dataset, task, and metric and select one primary metric for each dataset--task node.
Repeated observations are then aggregated, normalized, and oriented so that higher values indicate better performance.
The merged corpus contains 247,803 observed model--dataset performance pairs with values notated as $y_{dm}$, many fewer than the number of candidate models.
Appendix~\ref{sec:appendix-method-inputs} specifies metric selection, source resolution, and per-node sampling.

\noindent\textbf{Edges:  Typed relations connect evidence.}
The graph has model and dataset--task nodes.
Its five directed relations comprise observed performance and its reverse, dataset similarity, and parent-to-derivative lineage and its reverse.
Performance edges represent recorded evaluations, while similarity connects each dataset to its ten nearest neighbors by descriptor cosine and %
lineage connects parent and derivative model nodes using resolved parent-model references in the model metadata.
These relations connect sparse observations with metadata and other structural evidence without treating lineage or similarity as measured performance.

\subsection{Relation-Aware Representation Learning}
\label{sec:graph-encoder}

\noindent\textbf{Metadata anchors both node types.}
Model and dataset--task inputs combine a name hash and a text embedding from a %
sentence transformer
~\citep{wang2020minilm} with learned type-specific lookup features.
Model text describes the repository name, family, and parameter size when known; dataset text uses a matched card when available and otherwise uses its name and task.
Dataset nodes also include root-level statistics.
The resulting model and dataset--task inputs have 480 and 486 dimensions, respectively.
Node inputs contain neither performance statistics nor gold-eligibility indicators; feature construction is detailed in Appendix~\ref{sec:appendix-method-inputs}.

\begin{figure}[!htbp]
  \centering
  \includegraphics[width=\linewidth]{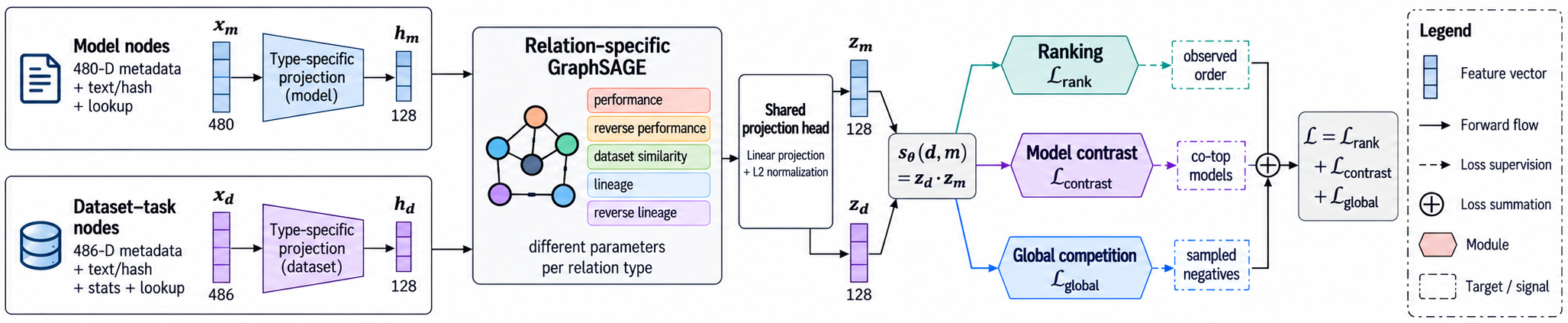}
  \caption{Representation learning and training.
  Model and dataset--task features pass through type-specific projections, a relation-specific GraphSAGE encoder, and a shared projection head.
  Ranking and global losses compare query--model pairs; model contrast compares model pairs.}
  \label{fig:embedding-construction}
\end{figure}

\noindent\textbf{Relation-aware aggregation yields indexable vectors.}
Indexable retrieval requires model vectors that do not depend on the incoming query.
We use a relation-specific GraphSAGE layer~\citep{hamilton2017graphsage} to independently compute model and query representations.
Type-specific projections map both node types into a shared hidden space, followed by relation-aware neighborhood aggregation and a shared output head.
The resulting unit-length representations are scored by inner product,
$s_\theta(d,m)=z_d^\top z_m$.
The detailed encoder formulation is provided in Appendix~\ref{app:graph-encoder}.
The key property of this architecture is that model and query representations are computed independently. This means model vectors can be indexed once offline and searched for any query without rescoring the full lake. The design achieves this by sharing a single output head across both node types, placing models and queries in the same vector space, while keeping aggregation unweighted.  Performance values provide supervision  and dataset similarity determines graph topology, but neither scales the messages passed between neighbours. Hash and text features remain frozen; lookup tables, projections, gates, and the shared head are learned.
\FloatBarrier

\subsection{Training with Held-Out Evidence}
\label{sec:learning-objectives}

\noindent\textbf{Root splits control evidence visibility.}
A dataset root groups configurations of the same dataset. 
All nodes sharing a dataset root belong to the same training, validation, or test partition, so related configurations do not cross evaluation boundaries.
A disjoint portion of training performance edges supplies supervision; the remaining training edges carry messages.
Let $E_{\mathrm{tr}}$ denote performance edges from training roots.
Validation message passing can use all training edges, and test message passing can also use validation edges, but neither graph contains the performance edges of the roots it evaluates.
Reverse performance edges follow the same rule, while dataset similarity and lineage remain fixed.
Reported indexes and query vectors use the test message graph, making this transductive retrieval with held-out performance evidence.
The exact edge sets and split proportions appear in Appendix~\ref{sec:appendix-training-details}.

\noindent\textbf{Training aligns the retrieval score.}
We jointly optimize local ranking, model contrast, and whole-lake competition with equal weight.
\begin{equation}
\mathcal L=\mathcal L_{\mathrm{rank}}+\mathcal L_{\mathrm{contrast}}+\mathcal L_{\mathrm{global}}
\label{eq:total-objective}
\end{equation}
All three objectives use the unit-normalized representations in Eq.~\ref{eq:graph-retrieval-encoder}.
Their temperatures $\tau_{\mathrm{rank}}$, $\tau_{\mathrm{contrast}}$, and $\tau_{\mathrm{global}}$ are listed in Appendix~\ref{sec:appendix-training-details}.

\noindent\textbf{Local ranking preserves observed order.}
For each supervised query $d$, we retain preference pairs $\mathcal P_d\subseteq\{(i,j):y_{di}-y_{dj}>0.02\}$.
RankNet~\citep{burges2005ranknet} penalizes retrieval scores that reverse their observed performance order.
\begin{equation}
 \mathcal L_{\mathrm{rank}}=
 \frac{1}{|\mathcal D_B|}\sum_{d\in\mathcal D_B}
 \frac{1}{|\mathcal P_d|}\sum_{(i,j)\in\mathcal P_d}
 \operatorname{softplus}\!\left(-\frac{s_\theta(d,i)-s_\theta(d,j)}{\tau_{\mathrm{rank}}}\right)
 \label{eq:local-ranknet}
\end{equation}
Here $\mathcal D_B$ contains batch queries with retained pairs.
We keep at most 256 pairs per query; when capped, half are the most inverted under the current model and half are sampled from the remainder.
Averaging within queries limits the influence of densely observed datasets.

\noindent\textbf{Model contrast links strong models.}
For a query $d$ with $n_d$ training-visible observations, let $H_d$ contain its top $k_d=\max(1,\operatorname{round}(0.1n_d))$ models.
Distinct models sharing membership in any $H_d$ form positive pairs.
Let $\mathcal P_i$ be the batch positives for anchor $i$, and let $\mathcal A$ contain anchors with nonempty $\mathcal P_i$.
For each anchor, we draw 256 batch models uniformly with replacement and remove self IDs and positives, leaving a negative multiset $\mathcal N_i$.
With $\phi(i,j)=z_i^\top z_j/\tau_{\mathrm{contrast}}$, the contrastive objective is
\begin{equation}
 \mathcal L_{\mathrm{contrast}}=\frac{1}{|\mathcal A|}\sum_{i\in\mathcal A}
 \left[\log\!\left(\sum_{p\in\mathcal P_i}e^{\phi(i,p)}+
 \sum_{n\in\mathcal N_i}w_{in}e^{\phi(i,n)}\right)
 -\frac{1}{|\mathcal P_i|}\sum_{p\in\mathcal P_i}\phi(i,p)\right]
 \label{eq:model-contrast}
\end{equation}
The negative weight $w_{in}$ is two for models in the anchor's lineage component and one otherwise: shared ancestry alone does not imply similar performance.

\noindent\textbf{Global competition covers the lake.}
Local batches expose each query to only a small portion of the model lake.
We draw additional negatives from $q(m)=\tfrac12q_{\deg}(m)+\tfrac12q_{\mathrm{lab}}(m)$, where $q_{\deg}(m)\propto(\deg(m)+1)^{0.75}$ over all models and $q_{\mathrm{lab}}$ is uniform over models with positive training degree.
Degree counts observations in $E_{\mathrm{tr}}$ only.
Each step samples 128 positive-bearing queries and 256 negatives per query with replacement; removing members of $H_d$ leaves the multiset $S_d$.
Let $a_{dm}=s_\theta(d,m)/\tau_{\mathrm{global}}$ and let $\mathcal Q_B$ denote the sampled queries.
The sampled-softmax objective is
\begin{equation}
\mathcal L_{\mathrm{global}}
=
\frac{1}{|\mathcal Q_B|}
\sum_{d\in\mathcal Q_B}
\left[
\log\!\left(
\sum_{p\in H_d}e^{a_{dp}}
+
\sum_{n\in S_d}e^{a_{dn}-\log q(n)}
\right)
-
\frac{1}{|H_d|}\sum_{p\in H_d}a_{dp}
\right]
\label{eq:global-softmax}
\end{equation}
The $-\log q(n)$ correction applies only to sampled negatives.
This objective uses a fresh full-message-graph forward pass with edge dropout.
Mini-batches are rooted at disjoint supervision edges, whose labels are never dropped.
Optimization settings and split seeds are in Appendix~\ref{sec:appendix-training-details}.

\subsection{Indexed Retrieval and Reranking}
\label{sec:two-stage-retrieval}

\noindent\textbf{Indexed Candidate Retrieval.}
The score in Eq.~\ref{eq:graph-retrieval-encoder} uses model vectors that are independent of the query, so we add every model vector to an inner-product HNSW index~\citep{DBLP:journals/pami/MalkovY20} offline.
Given query vector $z_d$, the index retrieves $K=10^3$ nearby models from the full lake, forming the candidate pool $\widehat{\mathcal P}_K(d)$.
Only models in this pool proceed to task-prior reranking.
HNSW is an approximate search algorithm whose thoroughness is controlled by a parameter $ef\_search$ where larger values explore more of the index and find neighbours closer to the true nearest neighbours, at the cost of additional search time. We set $ef$ by comparing HNSW's retrieved candidates against exact brute-force nearest neighbours — finding the smallest $ef\_search$ that recovers nearly all exact neighbours — rather than tuning it directly against the gold evaluation labels.  Details are in Appendix~\ref{sec:appendix-search-calibration}.

\noindent\textbf{Historical Evidence-Based Reranker.}
For task $t$ and model $m$, let $n_{tm}$ and $A_{tm}$ be the count and sum of normalized evidence values on training and validation roots.
Canonical task strings define these groups; the scored test root and its sibling configurations are excluded.
We use a fixed shrinkage prior and fuse it with the normalized retrieval score.
Unseen task--model pairs receive no historical bonus, and no query-wise min--max rescaling is applied.
Only retrieved candidates receive prior lookups; a fixed label-free model order breaks exact ties before returning the top ten.
This reranker is deterministic and cannot recover a model omitted from the candidate pool.
The exact prior and score-fusion formulation is given in Appendix~\ref{sec:appendix-reranker}.

%% file: src/experiment.tex
\section{Experiments}\label{sec:exp}

For each test dataset and task, we recommend ten models. We check whether
the highest-scoring model in the available test records is among those
ten. We also measure how long it takes to search the model lake. We
repeat the test using three different train/validation/test splits. The
two tables below report results for one of these splits.
Appendix~\ref{sec:appendix-detailed-results} gives more details.

\subsection{Evaluation Protocol}
\label{sec:eval-protocol}

\paragraph{Dataset: Evaluation queries and data splits.}
The lake is sparse: at least 92\% of the 3,016,439 candidate models have no recorded performance pair, and model size is unknown for 72\% of candidates. This sparsity is the central challenge that ModelLakeFishing is designed to address.
Each query names a dataset and a task already in the graph. A dataset
root groups related configurations of one dataset. We divide the data
into training, validation, and test groups in three different ways and
keep every configuration of a dataset in the same group. We retrain
the model for each split using the same initialization seed. For a test
query, the model can see the dataset, task, and their metadata, but it
cannot see the links that record how models performed on that dataset.
We remove those links in both graph directions and use their scores only
to check the returned list. The task prior also uses no results from
the test dataset or its related versions. 
The corpus is constructed as follows.
The snapshot has 1,008,417 dataset records. We make graph nodes only for
dataset--task combinations with recorded evaluations. After normalizing
dataset names, grouping records by dataset and task, and choosing one
metric for each group, we have 18,729 dataset--task nodes. Removing
duplicate or conflicting records and limiting records per node leaves
247,803 model--dataset performance pairs. Dataset cards match 3,928
nodes (21\%).
For testing, we exclude
reinforcement-learning tasks and placeholder dataset names, because reinforcement-learning reward metrics have no consistent orientation, and placeholder names indicate unreliable dataset provenance.
Each test
query must have at least three models with recorded scores that are not
all equal. We must also know whether higher or lower is better for the
metric. The three different splits contain 1,476, 1,101, and 1,545 test
queries from 730, 495, and 473 dataset roots, respectively.
The splits differ in size because the 70/10/20 edge-proportion target is applied to dataset roots of varying size, producing unequal test query counts.
Appendix~\ref{sec:appendix-data} lists
the six input sources. Appendices~\ref{sec:appendix-method-inputs}
and~\ref{sec:appendix-training-details} give the construction and
training settings.

\paragraph{Baselines.}
We compare three search methods using the same learned vectors and task
prior. The task prior uses a model's past results on other datasets with
the same task. The first method (\exhaustive) scores and 
ranks all 3,016,439 models.
The second checks similarity to every model (\exact), keeps the closest 1k,
and reranks that smaller set. The third (\ours) uses the HNSW index to find
1k nearby models and reranks them. Comparing the first two shows the
effect of limiting the candidate set; comparing the last two shows the
extra effect of approximate nearest-neighbour search. LEEP, LogME, NLEEP, and H-score
compute a score from the target data for each candidate. TransferGraph
and ModelLens also score each query--candidate pair.  %
However, these approaches do not scale to the massive data lake we consider here.

\paragraph{Lookup time.}
We start the timer after the query vector has been computed. We stop it
after HNSW search, task-prior lookup, combining the scores, and sorting.
The time does not include computing the query vector, loading the index,
or sending data over a network. The reported full-lake scan was timed
over a different set of steps, so its time cannot be used to calculate
an online speedup. Appendix~\ref{sec:appendix-cost-accounting} gives
the other recorded times and memory measurements.

\subsection{Experimental Results}
\label{sec:experimental-results}

\begin{figure}[!htbp]
 \centering
 \includegraphics[width=\linewidth]{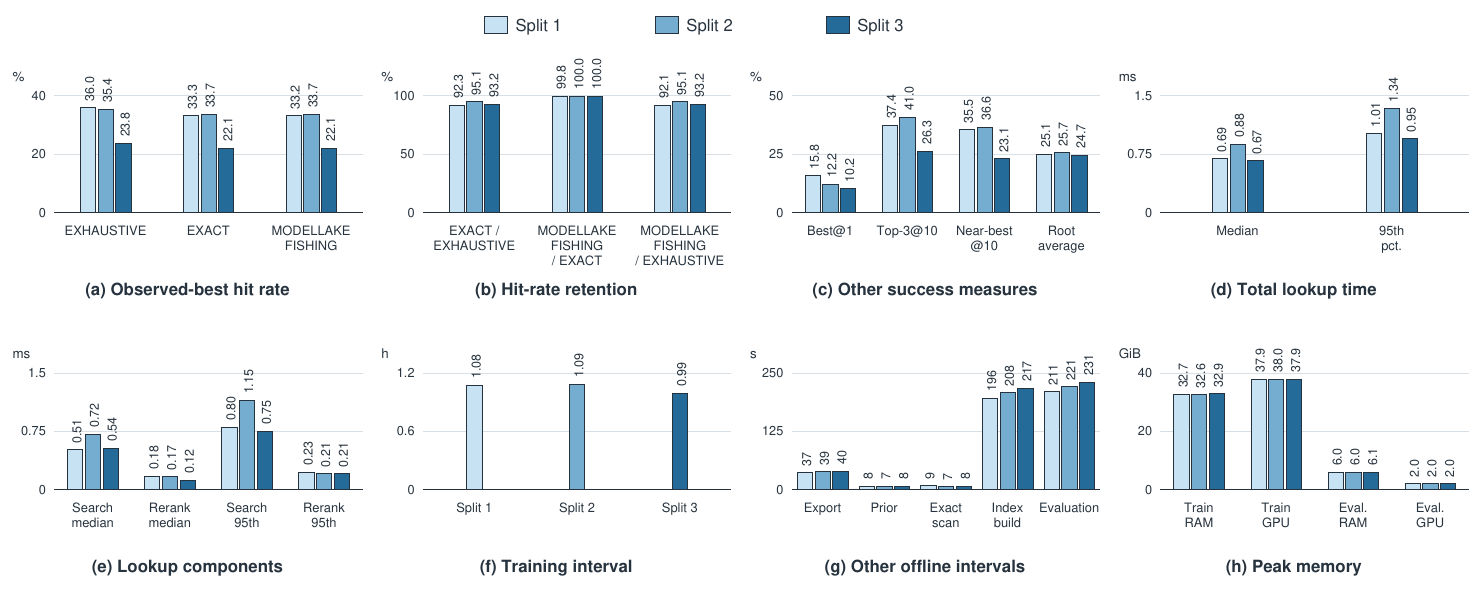}
 \caption{Results for three dataset-root splits. Within each group, bars
 follow the legend from left to right; numbers above bars are rounded.} 
 \label{fig:a0-retrieval-results}
\end{figure}

For each test query $d$, let $g_d$ be the highest-scoring model among
those with recorded test results. We first adjust each metric so that
higher scores are always better. A query is a \emph{hit at \(k\)} when
$g_d$ appears among the first \(k\) returned models. Gold@\(k\) is the
share of test queries that are hits.
\begin{equation}
 \bigl(\texttt{gold@}k\bigr)_s=
 \frac{1}{|\mathcal Q_s|}\sum_{d\in\mathcal Q_s}
 \mathbf{1}[g_d\in R_k(d)]
 \label{eq:gold-at-k}
\end{equation}
We mainly report \(k=10\) and average the three split results with equal
weight. This measure checks the best model we have tested, not whether
an untested model might be better. Appendix~\ref{sec:appendix-eval-metrics}
defines the other measures in Figure~\ref{fig:a0-retrieval-results}(d).
The results reported in the  two tables below are based on a split with 1,476 queries. MRR@10 averages 1 divided by the
rank of the best tested model when it appears in the top ten; it reports
zero otherwise. Gold coverage (Gold cov.) at 1k is the share of queries
for which the best tested model enters the 1k candidate pool before
reranking.

\begin{table}[!htbp]
 \centering
 \footnotesize
 \setlength{\tabcolsep}{3pt}
 \renewcommand{\arraystretch}{1.15}
 \resizebox{\linewidth}{!}{%
 \begin{tabular}{@{}lrrrrr@{}}
 \toprule
\textbf{Run}\footnotemark & \textbf{Gold@1} & \textbf{Gold@5} & \textbf{Gold@10} & \textbf{MRR@10} & \textbf{Gold cov.@1k} \\
 \midrule
 \multicolumn{6}{@{}l@{}}{\textit{Ablation results : w/ task prior}} \\
 \midrule
\ours (full) & $\mathbf{15.8537}$ & $29.6070$ & $33.1978$ & $21.7172$ & $53.9295$ \\
w/o graph encoder & $13.4146_{\textcolor{red!70!black}{\downarrow2.44}}$ & $27.3713_{\textcolor{red!70!black}{\downarrow2.24}}$ & $29.1328_{\textcolor{red!70!black}{\downarrow4.07}}$ & $19.3052_{\textcolor{red!70!black}{\downarrow2.41}}$ & $39.4986_{\textcolor{red!70!black}{\downarrow14.43}}$ \\
w/o ranking & $15.5827_{\textcolor{red!70!black}{\downarrow0.27}}$ & $\mathbf{30.0136}_{\textcolor{teal!80!black}{\uparrow0.41}}$ & $\mathbf{35.0271}_{\textcolor{teal!80!black}{\uparrow1.83}}$ & $\mathbf{21.8393}_{\textcolor{teal!80!black}{\uparrow0.12}}$ & $\mathbf{58.6721}_{\textcolor{teal!80!black}{\uparrow4.74}}$ \\
w/o model contrast & $14.7019_{\textcolor{red!70!black}{\downarrow1.15}}$ & $26.6260_{\textcolor{red!70!black}{\downarrow2.98}}$ & $29.8103_{\textcolor{red!70!black}{\downarrow3.39}}$ & $19.8643_{\textcolor{red!70!black}{\downarrow1.85}}$ & $47.9675_{\textcolor{red!70!black}{\downarrow5.96}}$ \\
w/o global competition & $0.0000_{\textcolor{red!70!black}{\downarrow15.85}}$ & $0.0000_{\textcolor{red!70!black}{\downarrow29.61}}$ & $0.0000_{\textcolor{red!70!black}{\downarrow33.20}}$ & $0.0000_{\textcolor{red!70!black}{\downarrow21.72}}$ & $0.0000_{\textcolor{red!70!black}{\downarrow53.93}}$ \\
 \midrule
 \multicolumn{6}{@{}l@{}}{\textit{Ablation results : wo/ task prior}} \\
 \midrule
\ours (full) & $\mathbf{1.8970}$ & $6.7751$ & $12.0596$ & $\mathbf{4.3722}$ & $53.9295$ \\
w/o graph encoder & $1.3550_{\textcolor{red!70!black}{\downarrow0.54}}$ & $\mathbf{7.7236}_{\textcolor{teal!80!black}{\uparrow0.95}}$ & $11.6531_{\textcolor{red!70!black}{\downarrow0.41}}$ & $4.0667_{\textcolor{red!70!black}{\downarrow0.31}}$ & $39.4986_{\textcolor{red!70!black}{\downarrow14.43}}$ \\
w/o ranking & $1.3550_{\textcolor{red!70!black}{\downarrow0.54}}$ & $7.1816_{\textcolor{teal!80!black}{\uparrow0.41}}$ & $\mathbf{13.2114}_{\textcolor{teal!80!black}{\uparrow1.15}}$ & $4.0573_{\textcolor{red!70!black}{\downarrow0.31}}$ & $\mathbf{58.6721}_{\textcolor{teal!80!black}{\uparrow4.74}}$ \\
w/o model contrast & $1.2195_{\textcolor{red!70!black}{\downarrow0.68}}$ & $3.2520_{\textcolor{red!70!black}{\downarrow3.52}}$ & $5.5556_{\textcolor{red!70!black}{\downarrow6.50}}$ & $2.1613_{\textcolor{red!70!black}{\downarrow2.21}}$ & $47.9675_{\textcolor{red!70!black}{\downarrow5.96}}$ \\
w/o global competition & $0.0000_{\textcolor{red!70!black}{\downarrow1.90}}$ & $0.0000_{\textcolor{red!70!black}{\downarrow6.78}}$ & $0.0000_{\textcolor{red!70!black}{\downarrow12.06}}$ & $0.0000_{\textcolor{red!70!black}{\downarrow4.37}}$ & $0.0000_{\textcolor{red!70!black}{\downarrow53.93}}$ \\

\midrule
 \multicolumn{6}{@{}l@{}}{\textit{Baseline comparison}} \\
 \midrule
Full model + task prior & $\mathbf{15.8537}$ & $\mathbf{29.6070}$ & $\mathbf{33.2656}$ & $\mathbf{21.7284}$ & $\mathbf{53.8618}$ \\
Full model w/o task prior & $1.8970_{\textcolor{red!70!black}{\downarrow13.96}}$ & $6.7751_{\textcolor{red!70!black}{\downarrow22.83}}$ & $12.0596_{\textcolor{red!70!black}{\downarrow21.21}}$ & $4.3722_{\textcolor{red!70!black}{\downarrow17.36}}$ & $\mathbf{53.8618}_{=}$ \\
Frozen MiniLM & $0.9485_{\textcolor{red!70!black}{\downarrow14.91}}$ & $2.4390_{\textcolor{red!70!black}{\downarrow27.17}}$ & $3.9973_{\textcolor{red!70!black}{\downarrow29.27}}$ & $1.6166_{\textcolor{red!70!black}{\downarrow20.11}}$ & $15.1084_{\textcolor{red!70!black}{\downarrow38.75}}$ \\
Popularity & $0.0000_{\textcolor{red!70!black}{\downarrow15.85}}$ & $0.0000_{\textcolor{red!70!black}{\downarrow29.61}}$ & $0.0000_{\textcolor{red!70!black}{\downarrow33.27}}$ & $0.0000_{\textcolor{red!70!black}{\downarrow21.73}}$ & $0.9485_{\textcolor{red!70!black}{\downarrow52.91}}$ \\
FM (metadata-adapted) & $0.0000_{\textcolor{red!70!black}{\downarrow15.85}}$ & $0.0000_{\textcolor{red!70!black}{\downarrow29.61}}$ & $0.0000_{\textcolor{red!70!black}{\downarrow33.27}}$ & $0.0000_{\textcolor{red!70!black}{\downarrow21.73}}$ & $0.1355_{\textcolor{red!70!black}{\downarrow53.73}}$ \\
 \bottomrule
 \end{tabular}%
 }
\end{table}
\footnotetext{Values are percentages. Subscripts show changes relative to the first row. The three methods use no task prior. FM is a metadata adaptation fitted on 198,216 training and validation observations; the full encoder is fitted on the training partition.}

\noindent\textbf{Main result}
On the seed-0 split, the full HNSW system achieves 33.20\% gold@10 and 53.93\% gold coverage@1k (first table).
Removing the task prior leaves pool coverage unchanged but lowers gold@10 to 12.06\%, showing that the prior improves ranking within the retrieved pool.
Removing the graph encoder or model contrast lowers gold@10 to 29.13\% and 29.81\%, respectively; removing global competition yields zero coverage on this split.
The ranking-loss ablation moves in the opposite direction, raising gold@10 to 35.03\%, so this split does not establish a benefit from that loss.
We compare three baselines without a task prior:
frozen MiniLM~\citep{wang2020minilm}, a pretrained text encoder
that ranks models by query--model cosine similarity;
popularity, which ranks by downloads; and a metadata-adapted
Factorization Machine (FM)~\citep{li2023guided}, which predicts
performance using pairwise metadata feature interactions.
In the exact top-1k comparison, the full model without the prior
achieves 12.06\% gold@10, versus 4.00\% for MiniLM and 0\% for
both popularity and FM. Adding the prior raises gold@10 to 33.27\%.

\noindent\textbf{Comparison to Baselines.}
Figure~\ref{fig:a0-retrieval-results} extends the retrieval comparison to three dataset-root splits and reports other evaluation views.
Panel (a) shows that (averaging over all splits), the mean gold@10 (best observed hit rate) changes from 31.74\% with \exhaustive
to 29.70\% with \exact
and 29.68\% with 
\ours.
This shows that almost all of the loss comes from limiting the pool, rather than from approximate index search.  Panel (b) reports the hit-rate retention for all three methods.
Panel (c) applies alternative success measures to the same returned lists (reported only for \ours).
Panels (d--e) report lookup time after query encoding, with a median of 0.747~ms and a 95th percentile of 1.102~ms.
Panels (f--h) report training, other processing intervals, and memory separately (time not included in the query time).

\FloatBarrier

%% file: src/conclusion.tex
\section{Conclusion}~\label{sec:conclusion}

Model lakes containing millions of reusable models present a retrieval challenge that exhaustive scoring cannot meet. Identifying suitable models for a new dataset requires learning from evidence that is sparse, heterogeneous, and unevenly distributed across a corpus too large to evaluate candidate by candidate. \ours addresses this by learning indexable representations from a model–dataset evidence graph.  We combine performance records, model cards, dataset descriptions, and model lineage into a structured graph.  We index embeddings with HNSW so that retrieval requires no candidate-wise execution at query time.

Evaluated on a lake of 3,016,439 models with a leakage-safe, root-aware protocol across three dataset partitions, \ours  achieves a mean gold@10 of 29.68\% — recovering the observed best model in the top ten for nearly 30\% of eligible queries. It retains 93\% of the gold@10 achieved by exhaustive scoring, with a median query latency of 0.747ms after query encoding. Nearly all of the quality gap relative to exhaustive search comes from limiting the candidate pool to 1,000 models; the HNSW approximation itself costs less than 0.1 percentage points of gold@10. The task-level prior is the dominant reranking signal: removing it drops gold@10 from 33\% to 12\% on the seed-0 split, while pool coverage remains unchanged, confirming that the prior improves ranking within the retrieved set rather than retrieval itself.

These results demonstrate that an ANN-searchable representation space over millions of models can be learned from relational supervision alone, without executing any candidate model on the target dataset. The evaluation measures recovery of historically observed models under sparse supervision; the true best model for a query may be unobserved and unreachable by any retrieval system operating over recorded evidence alone. Competitors do not scale to the size of graph we have considered, but a full comparison with methods that score individual query–candidate pairs, such as TransferGraph and ModelLens, on a matched candidate subset remains an important open experiment

%% file: src/appendix.tex
\section{Model Lake Visualization}
\label{sec:appendix-lake}

\begin{figure}[!htbp]
  \centering
  \includegraphics[width=0.90\textwidth,height=0.72\textheight,keepaspectratio]{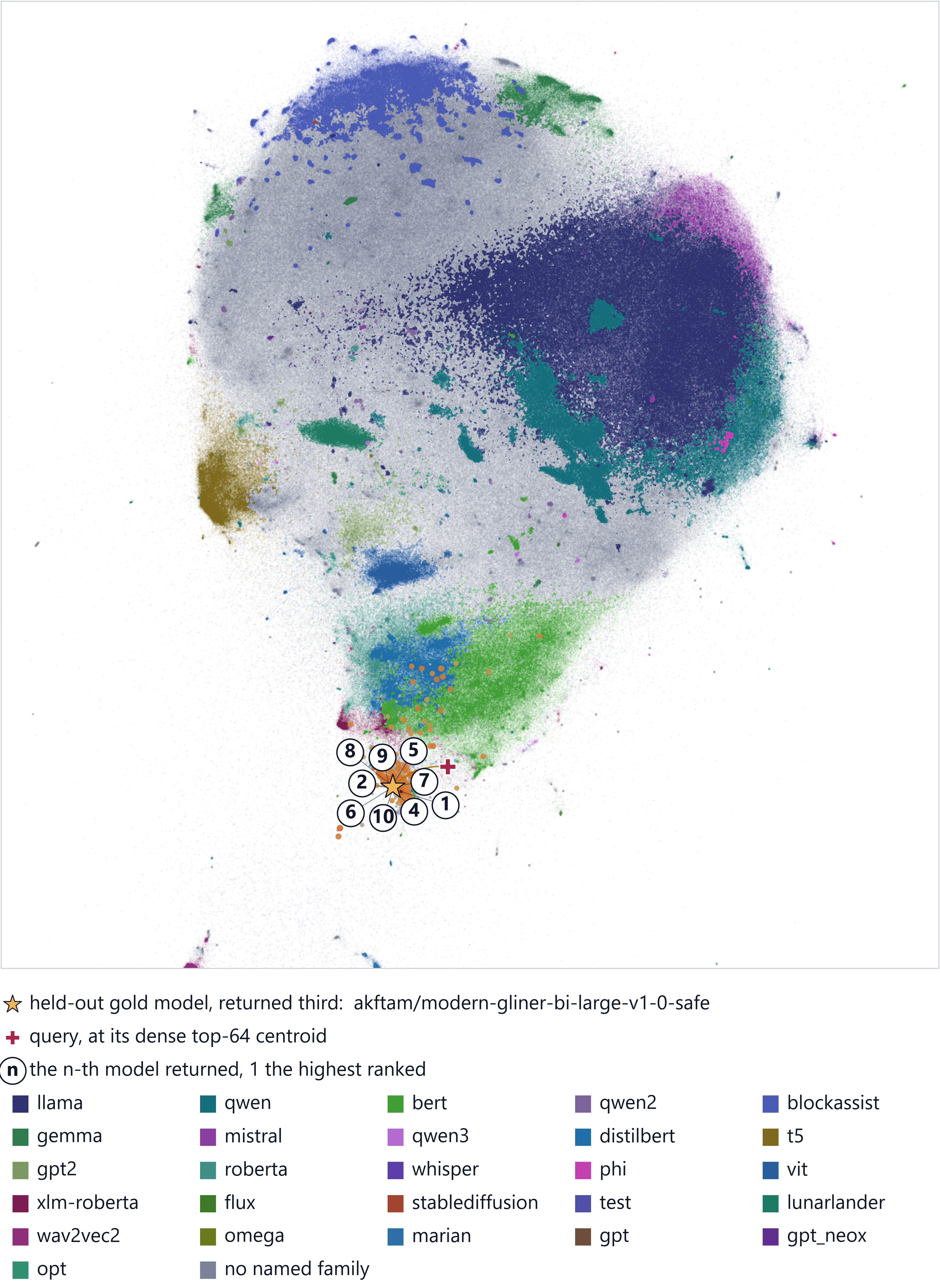}
  \caption{%
    \textbf{The model lake.} A UMAP projection of the $3{,}016{,}439$ learned
    model representations in the shared $128$-dimensional retrieval space,
    colored by model family; a small peripheral portion lies outside the
    plotted frame. The overlaid held-out \texttt{broad tweet
    corpus}\,/\,\emph{token-classification} example marks the query (plus
    sign at the centroid of its dense top-64 models), its gold model (star),
    and the ten returned models (numbered circles), showing how one
    retrieval is situated within the full lake.%
  }
  \label{fig:lake-ab}
\end{figure}

\section{Supplementary Method Details}
\label{sec:appendix-method}

\subsection{Evidence Sources}
\label{sec:appendix-data}

Table~\ref{tab:source-counts} records the contribution of each input
evaluation collection. These are data inputs, not baseline methods or
independent runs of the proposed retriever. The names are provenance IDs:
for example, \texttt{diverse\_zoo} denotes one historical evaluation
collection, not a method called ``Diverse Zoo.'' Its underlying
collection is not further characterized in the available evidence.

\begin{table}[htbp]
 \centering
 \caption{Retained model--dataset performance pairs by input source ID
 after deduplication, conflict resolution, and per-node capping.}
 \label{tab:source-counts}
 \small
 \begin{tabular}{lr}
  \hline
  Input source ID & Pairs \\
  \hline
  \texttt{modellens\_v2} & 117,898 \\
  \texttt{hf\_model\_index} & 73,670 \\
  \texttt{d0\_v1\_5} & 45,992 \\
  \texttt{hf\_effective} & 5,223 \\
  \texttt{a\_ctrl\_2000m} & 3,670 \\
  \texttt{diverse\_zoo} & 1,350 \\
  \hline
  Total & 247,803 \\
  \hline
 \end{tabular}
\end{table}

\input{src/appendix_method}
\input{src/appendix_evidence}

%% file: src/appendix_method.tex
\subsection{Evidence Corpus and Node Features}
\label{sec:appendix-method-inputs}

\paragraph{Construction of the evaluated lake.}
The frozen hub snapshot contains 3,003,759 model records and 1,008,417
dataset records. Historical evaluations contribute 12,680 additional
model IDs to the candidate universe. The parser reads 2,158,375 native
metric rows, retains 2,097,081 finite and parseable rows, and reduces
these to 1,435,162 after within-key median deduplication. Across the six
retained evidence sources, the merge starts from 531,958 rows, removes
5,102 within-source duplicates, resolves 1,502 cross-source conflicts,
and applies per-dataset capping. The
source breakdown is given in Table~\ref{tab:source-counts}; it describes
data provenance rather than competing retrieval methods.

\paragraph{Metric and feature processing.}
Native evaluation records are grouped by dataset, task, and full metric
name, retaining cutoff distinctions. Invalid numerical values are
discarded; repeated model observations are median-aggregated and
min--max normalized within each group, with constant groups assigned
0.5. Lower-is-better values are oriented as $y=1-\bar v$. Each node
selects one primary metric, preferring known direction, then reward,
then unknown; observation count and lexical order break ties.
Reward/unknown values (2.03\% of retained performance edges) remain
numerical evidence for learning and priors but cannot define evaluation
gold. Historical sources retain their stored orientation: duplicate
(source, node, model) records are median-aggregated and normalized
within each (source, node) group. Fixed source priority resolves
overlapping model--node pairs, and deterministic weight-stratified
sampling caps each node at 200 observations.

Node inputs combine a frozen 64-dimensional name hash with a
384-dimensional embedding from \texttt{all-MiniLM-L6-v2}. Model text
contains the cleaned repository name, canonical family, and a
parameter-size phrase when known. Dataset text contains its cleaned
name and task, matched card task categories, up to ten non-colon tags,
and at most 400 description characters when a card match is reliable.
Dataset nodes append a ten-slot statistics block whose only nonzero
coordinate is the seventh, $\log(1+r_d)$, where $r_d$ counts frozen
dataset--task nodes sharing the same root. Learned size and family
lookups add 16 dimensions each to model inputs, with explicit unknown
categories; task-type,
class-count, and arity lookups add 16, 8, and 4 dimensions to dataset
inputs. The resulting input dimensions are 480 and 486.

\begin{figure}[htbp]
 \centering
 \includegraphics[width=\linewidth]{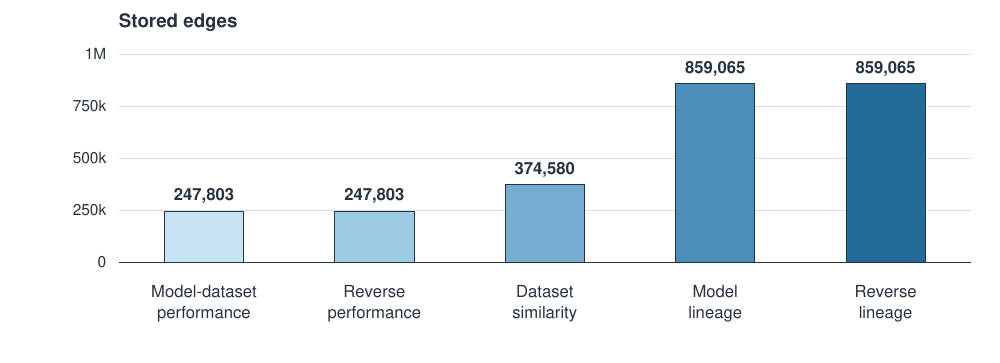}
 \caption{Stored directed edges by graph relation. Reverse directions
 are separate message relations. The five bars sum to 2,588,316 edges;
 the total is not itself a relation.}
 \label{fig:a0-relations}
\end{figure}

The model feature matrix has 448 columns. Dataset--task features have
458 columns; seven performance-derived input columns are zeroed in the
evaluated configuration. The graph relation counts appear in
Figure~\ref{fig:a0-relations}. For training, each dataset keeps its ten
nearest dataset neighbors from a stored top-20 graph, and those links
contribute equally; the stored
count reflects the larger construction graph. These are properties of the fixed input and
training recipe, not component-removal experiments.

\subsection{Training Configuration and Evidence Visibility}
\label{sec:appendix-training-details}

\paragraph{Fixed training configuration.}
The encoder uses one relation-specific GraphSAGE layer with 128-dimensional
output. Relation gates are initialized to one. Adam optimization runs
for 25 epochs at learning rate 0.01 with
supervision batches of 1,024. Ordinary and lineage message-edge dropout
rates are 0.30 and 0.05. The ranking, model-contrast, and lake-wide loss
terms have equal weight; their temperatures $\tau_{\mathrm{rank}}$,
$\tau_{\mathrm{contrast}}$, and $\tau_{\mathrm{global}}$ are 0.10, 0.20,
and 0.10, respectively. Model contrast draws 256 batch models per
anchor, while the global objective draws 256 negatives per query.
The final checkpoint is used; MSE, uniformity regularization,
separate projection heads, and early stopping are disabled. This
paragraph records the configuration, not an ablation.

\paragraph{Evidence visibility.}
Dataset roots are assigned to training, validation, and test, targeting
approximately 70/10/20\% of performance edges. Let the forward edge
sets be $E_{\mathrm{tr}},E_{\mathrm{val}},E_{\mathrm{te}}$. A disjoint
30\% of $E_{\mathrm{tr}}$ is reserved as supervision $E_{\mathrm{sup}}$.
The message graphs expose these performance edges:
\begin{equation}
 E^{\mathrm{msg}}_{\mathrm{train}}=E_{\mathrm{tr}}\setminus E_{\mathrm{sup}},
 \quad E^{\mathrm{msg}}_{\mathrm{val}}=E_{\mathrm{tr}},
 \quad E^{\mathrm{msg}}_{\mathrm{test}}=E_{\mathrm{tr}}\cup E_{\mathrm{val}}.
 \label{eq:message-visibility}
\end{equation}
Reverse performance edges follow the same visibility. Root split seeds
are 0, 1, and 2, with initialization seed fixed at 0.

\subsection{Historical Evidence-Based Reranker}
\label{sec:appendix-reranker}
For task $t$ and model $m$, let $n_{tm}$ denote the number of historical 
evidence values and $A_{tm}$ their sum over eligible training and validation 
roots. We use a shrinkage prior with strength $5$ and prior mean $0.5$:
\begin{equation}
 p_t(m)=\begin{cases}
 \dfrac{A_{tm}+5\cdot0.5}{n_{tm}+5},&n_{tm}>0,\\[3pt]
 0,&n_{tm}=0,
 \end{cases}
 \qquad r(d,m)=\frac{s_\theta(d,m)+1}{2}+p_t(m)
 \label{eq:task-prior-fusion}
\end{equation}
Here $(s_\theta(d,m)+1)/2$ maps the cosine-style retrieval score to 
$[0,1]$, while $p_t(m)$ provides the historical task-specific bonus. 
For unseen task--model pairs, $p_t(m)=0$.

\subsection{Relation-Aware Graph Retrieval Encoder}
\label{app:graph-encoder}
Type-specific affine maps project both node types to
$h_v^{(0)}\in\mathbb R^{128}$.
For each relation $r$, let $\mathcal N_r(v)$ be the neighbors of node $v$
under that relation, and let $\mathcal R(v)$ be the relation types whose
destination is the type of $v$.

One relation-specific GraphSAGE layer~\citep{hamilton2017graphsage}
first averages these neighbors, then combines their features with the
node's own features using parameters for that relation:
\begin{equation}
 \begin{alignedat}{2}
 \mu_{v,r}&=\frac{\sum_{u\in\mathcal N_r(v)}h_u^{(0)}}{\max\{1,|\mathcal N_r(v)|\}},\qquad&
 h_v^{(1)}&=\sum_{r\in\mathcal R(v)}g_r
  \bigl(W_{s,r}h_v^{(0)}+W_{n,r}\mu_{v,r}+b_r\bigr),\\
 z_v&=\frac{W_Oh_v^{(1)}+b_O}{\|W_Oh_v^{(1)}+b_O\|_2},\qquad&
 s_\theta(d,m)&=z_d^\top z_m.
 \end{alignedat}
 \label{eq:graph-retrieval-encoder}
\end{equation}

Here $\mu_{v,r}$ is the average neighbor vector for relation $r$ and is zero
when $\mathcal N_r(v)$ is empty.
The sum for $h_v^{(1)}$ includes only relations in $\mathcal R(v)$;
$W_{s,r}$ and $W_{n,r}$ are the relation-specific self and neighbor
projections, and $g_r$ is a learned scalar gate.

%% file: src/appendix_evidence.tex
\section{Supplementary Evaluation Details}
\label{sec:appendix-detailed-results}

This section expands the protocol in Section~\ref{sec:eval-protocol} and
the measurements in Section~\ref{sec:experimental-results}. It records
the interpretation of Figure~\ref{fig:a0-retrieval-results};
it does not report additional comparison runs or ablations.

\subsection{Held-Out Query Eligibility}
\label{sec:appendix-query-eligibility}

Eligible dataset--task queries have a known metric direction or a
curated oriented source, exclude reinforcement-learning tasks and
placeholder dataset names, and have at least three nonconstant
observed held-out candidate values. The query counts in
Section~\ref{sec:eval-protocol} are observations across three root-aware
splits; the same node may occur in more than one split.
All queried nodes already exist in the frozen graph, so this is a
transductive historical-recovery evaluation, not a test of new-dataset
ingestion.

\subsection{Additional Recovery Metrics}
\label{sec:appendix-eval-metrics}

Figure~\ref{fig:a0-retrieval-results}(d) shows several readouts of the
same returned model lists. \texttt{gold@1} requires the observed best
model to rank first; \texttt{top3@10} accepts any of the three best
observed models among the returned ten; \texttt{gold-gap@10} accepts an
observed model within 0.01 of the best oriented performance. The
root-balanced measure first averages queries within each dataset root
and then averages roots. The median reranked gold position, conditional
on gold entering the candidate pool, is 4, 5, and 4 across the splits.
The frozen candidate order resolves ties in the observed-best label;
the same fixed model-ID order resolves ties in diagnostic scoring.

\subsection{Approximate-Search Calibration}
\label{sec:appendix-search-calibration}

The index uses inner-product search with HNSW construction parameters
$M=32$ and \texttt{ef\_construction}=200, using eight construction threads.
Search breadth is selected from 1k, 1,500, 2,000, 3,000, and 5,000:
the first setting with mean top-1k neighbor-ID recall of at least
0.99 against exact dense search is retained. This calibration does not
read gold labels. The selected \texttt{ef\_search} values for the three
splits are 1k, 1,500, and 1k; the corresponding neighbor-ID recall
values are 0.9944, 0.9963, and 0.9928. The task-prior tables are built from
training and validation roots only; the test root and its siblings are
excluded. This calibration checks search fidelity, not the ranking
benefit of the prior or of graph learning.

\subsection{Timing and Resource Accounting}
\label{sec:appendix-cost-accounting}

The formal evaluation ran on Linux with \texttt{hnswlib}~0.8.0. HNSW
search and prior fusion used the host CPU; exact diagnostic vector
scoring used an NVIDIA L40S. Each split has one timed query pass using
one HNSW query thread after up to 50 warm-up queries. The index was
built separately with eight threads; its recorded build interval includes
initialization, insertion, and serialization.

Figure~\ref{fig:a0-retrieval-results}(f) separates search from prior
lookup, fusion, and ordering. The total is timed per query before
percentiles are computed; component percentiles must not be added to
reconstruct a total percentile. All times exclude query encoding,
loading, and network overhead.

Figure~\ref{fig:a0-retrieval-results}(g--h) uses separate zero-based
scales for training and shorter intervals. The two exact references
share a scan, and the recorded evaluation interval already includes
index construction; these bars are not additive pipeline stages. Each
HNSW index occupies about 2.215~GiB. Training, vector export, and
formal evaluation ran on different hosts, so the intervals do not
form one same-machine runtime comparison.

Figure~\ref{fig:a0-retrieval-results}(i) separates training and
evaluation resources. Process RSS high-water marks and PyTorch GPU
allocated-memory peaks are different measures, not additive components.
The RSS highs can include allocations made earlier by the process.